\documentclass[aps,prl,twocolumn,groupedaddress]{revtex4}
\usepackage{epsfig}
\usepackage{psfig}
\usepackage{graphicx}
\date{\today}
\newcommand{\tkp}{\tilde{k}_\perp}
\newcommand{\bk}{{\bf k}}

\begin{document}
\title{Heat transport in ultra-thin dielectric membranes and bridges}
\author{T. K{\"u}hn$^1$, D. V. Anghel$^2$, J. P. Pekola$^3$, M. Manninen$^1$, and Y. M Galperin$^2$ \\ 
$^1$ University of Jyv\"askyl\"a, Department of Physics, P.O. Box 35 (YFL), FIN-40014 Jyv\"askyl\"a, Finland. \\
$^2$ University of Oslo, Department of Physics, P.O. Box 1048 - Blindern, N-0316 Oslo, Norway. \\
$^3$ Low Temperature Laboratory, Helsinki University of Technology, PO Box 2200, FIN - 02015 HUT, Finland
}
\begin{abstract}\noindent
Phonon modes and their dispersion relations in ultrathin homogenous
dielectric membranes are calculated using elasticity theory. The
approach differs from the previous ones by a rigorous account of the effect
of the film surfaces on the modes with different polarizations.
We compute the heat capacity of membranes
and the heat conductivity of narrow bridges cut out of such membranes, 
in a temperature range where the dimensions have a strong 
influence on the results. In the high temperature regime we recover 
the three-dimensional bulk results. However, in the low temperature 
limit the heat capacity, $C_V$, is proportional with $T$ (temperature), while 
the heat conductivity, $\kappa$, of narrow bridges is proportional to 
$T^{3/2}$, leading to a thermal cut-off frequency 
$f_c=\kappa/C_V\propto T^{1/2}$. 

\end{abstract}
\maketitle

\section{Introduction} \label{motiv}

More and more precise measurements are needed in science (e.g. in 
astrophysics) and therefore very sensitive detectors are required. 
However, even if all the 
technological difficulties are removed and the detector components
would be perfect, the thermal fluctuations and the discreteness of
particle and energy fluxes through the detector would  still limit
the precision of the measurements.
To achieve the desired sensitivity, the detectors have to work at sub-kelvin 
temperatures (to minimize the fluctuations) and have linear dimensions in the 
micrometer scale or even smaller (for small 
heat capacity and for the possibility to assamble high resolution 
detector arrays for space applications). 
For such ranges of temperatures and dimensions, finite-size effects 
come into play. 

\begin{figure}[ht]
\begin{center}
\unitlength1mm\begin{picture}(80,26)
\put(0,0){\epsfig{file=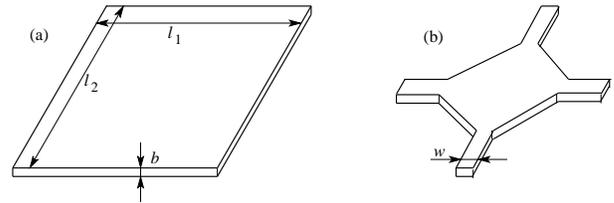,width=80mm}}
\end{picture}
\caption{(a) A schematic drawing of a thin free standing membrane of
the type discussed in this paper. The thickness, $b$, is typically a 
few hundreds of nm. The other dimensions, $l_1$ and $l_2$, are of the order 
of 100 ${\rm \mu m}$. (b) Such a membrane (central part) suspended 
to the bulk by four bridges. The bridge width, $w$, ranges from few 
${\rm \mu m}$ to tens of ${\rm \mu m}$.}\label{fig_membrane}
\end{center}
\end{figure}

Good thermal insulation of the detector is essential to attain the 
working temperature. For this purpose, placing the detector on top of a 
free-standing, amorphous and dielectric silicon-nitride (SiN$_x$) membrane 
appears to be a very convenient design (see for example Refs.
\cite{leivopekola,luukanen1,pekola}). The thickness of such a membrane 
is of the order of hundreds of nm, which is very small as compared to its 
other dimensions. The membrane may also be cut, so that the detector 
lies on its central and wider part, which is connected to the bulk 
material by narrow bridges (See Fig.~\ref{fig_membrane}).

Thermal characteristics of SiN$_x$ membranes and bridges have been 
investigated experimentally by several authors 
(as examples, see \cite{leivopekola,roukes1,richards,woodcraft}). 
SiN$_x$ is an amorphous dielectric and, if we neglect here the dynamic 
defects (see review \cite{dd} and references therein), its low 
temperature thermal properties are determined only by phonons.
At temperatures below a few degrees Kelvin, the 
bulk material (assumed to be isotropic) is well described by the Debye 
model, with typical values for the longitudinal and transversal sound 
velocities $c_t=6200$ m/s and $c_l=10300$ m/s, respectively \cite{wenzel}. 
Hence, the heat capacity can be expressed as $C_V^{3D}=K_1T^3$, 
where $T$ is the temperature and $K_1$ is a constant. 
If the mean free path of the phonons, $\ell$, 
is assumed to vary as $\ell\propto \lambda^{-s}$, where 
$\lambda$ is the phonon wavelength, the heat conductivity 
has the expression $\kappa^{3D}=K_2 T^{3-s}$, where again 
$K_2$ is a material dependent constant (see for example Ref. \cite{ziman}). 

Note that for the above given values of sound velocity, and at 
temperatures below 1 K, the bulk dominant phonon wavelength, 
$\lambda_{dom}\approx h c/(1.6 k_B T)$ (see \cite{ziman}, Chap. 8), is of 
several hundreds of nanometers. Thus it is comparable with the 
smallest dimensions of the structures measured in 
Refs. \cite{leivopekola,roukes1,richards,woodcraft}. Typically in these experiments, the measurements made at the lower end of 
the temperature interval showed qualitative differences from the 
measurements at the upper end of the interval, due to finite-size 
effects.

In this paper we will improve the model of Refs. \cite{dragos1,dragos2} 
by calculating rigorously the phonon modes and their dispersion relations 
in ultrathin membranes and narrow bridges. 
We shall show the thermodynamical consequences of the 
calculations. Such calculations appear to be well known to those working 
in the field of elasticity theory (see for example Refs. 
\cite{auld2,landau}), but apparently are not so familiar to the 
condensed matter community. Therefore, to 
make the paper more readable, we shall give some details of our 
calculations. To describe these effects, a simplified model was recently
used \cite{dragos1,dragos2}. Namely, it was assumed that the
phonon modes 
are  quantized along the direction perpendicular to the membrane
surface.  In this way, the three-dimensional (3D) (quasi)continuous
phonon spectrum  splits into what one could call two-dimensional (2D)
branches \cite{dragos1}. At low enough temperatures, only the lowest
branches are  populated by phonons and the membrane is described as a
2D isotropic  phonon gas. In this situation, if the mean free path of
the  phonons varies as $l\propto \lambda^{-s}$, the heat capacity
has the form $C_V^{2D}=K_3T^2$ and the thermal conductivity is
$\kappa^{2D}=K_4 T^{2-s}$, where $K_3$ and $K_4$ are material
dependent constants. This model describes the change of  the exponent
of the temperature dependence of heat conductivity,  as observed in
Refs.~\cite{leivopekola,roukes1,richards}, and could  fit relatively
well the experimental data \cite{dragos1}. On the other hand, the
observation that in narrow bridges the exponent of the temperature
dependence of the heat conductivity apparently converges to 1.5 as the
width of the  bridge decreases \cite{leivopekola,pekola}, as well as the
increase of the  ratio $\kappa/C_V$ with temperature
\cite{leivopekola}, cannot be  explained in this model.

\section{The model}

In Refs.~\cite{dragos1,dragos2} it was considered simply that the
longitudinal and transversal polarized phonon modes are independently
quantized by Neumann boundary conditions imposed at the membrane
surfaces. Nevertheless, in a rigorous analysis it must be taken into 
account that the modes with different polarization
couple at a free surface\cite{auld2} 
and because of this, the phonon modes in thin membranes are 
``distorted'' and show features that go beyond a simple treatment 
of the {\em quasi} 2D phonon gas.

For concreteness, we perform calculations for the practically
important case of  SiN$_x$ membranes
with thickness $100-200$nm having parallel surfaces.
Other dimensions of the  membranes are usually
of the  order of 100 ${\rm \mu m}$. Consequently, the membranes can be
considered as
infinitely long and wide. It is further assumed that the material is
isotropic. The
eigenmodes of these kind of systems are well known from acoustics  and
are called Lamb waves \cite{auld2}.

Similar to electricity theory, the physical acoustic fields can be expressed 
by a scalar and a vector potential, $\Phi$ and $\vec{\Psi}$. The
velocity fields of the longitudinal $l$ and transversal $t$ phonon modes are
then defined as $\vec{v}_l=\vec{\nabla}\cdot\Phi$ and 
$\vec{v}_t=\vec{\nabla}\times\vec{\Psi}$. Simple wave equations can be 
derived for the potentials, 
$\triangle\Phi=c_l^{-2}\partial_t^2\Phi$ and $\triangle\vec{\Psi}=c_t^{-2}
\partial_t^2\vec{\Psi}$. 
Here $c_l$
and $c_t (< c_l)$ are the longitudinal and transversal sound velocities,
respectively. In an infinite  three-dimensional space
this yields one longitudinal and two transversal plane wave solutions.

\begin{figure}[t]
\begin{center}
\unitlength1mm\begin{picture}(80,27)(0,0)
\put(0,0){\epsfig{file=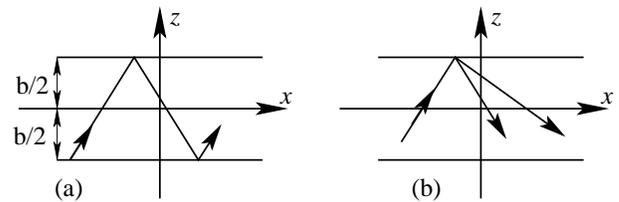,width=80mm}}
\end{picture}
\caption{Reflection of elastic waves at the free surfaces of the membrane 
of thickness $b$. 
A horizontal shear wave reflects also as a horizontal shear wave (a), while 
a wave longitudinally or transversally polarized into the $xz$ plane
reflects as a superposition of longitudinally and transversally polarized 
waves (b).}\label{reflections}
\end{center}
\end{figure}
\begin{figure*}[ht]
\centering
\setlength{\unitlength}{1.2 mm}\begin{picture}(120,30)
\put(-20,35){\includegraphics[width=40mm,angle=-90]{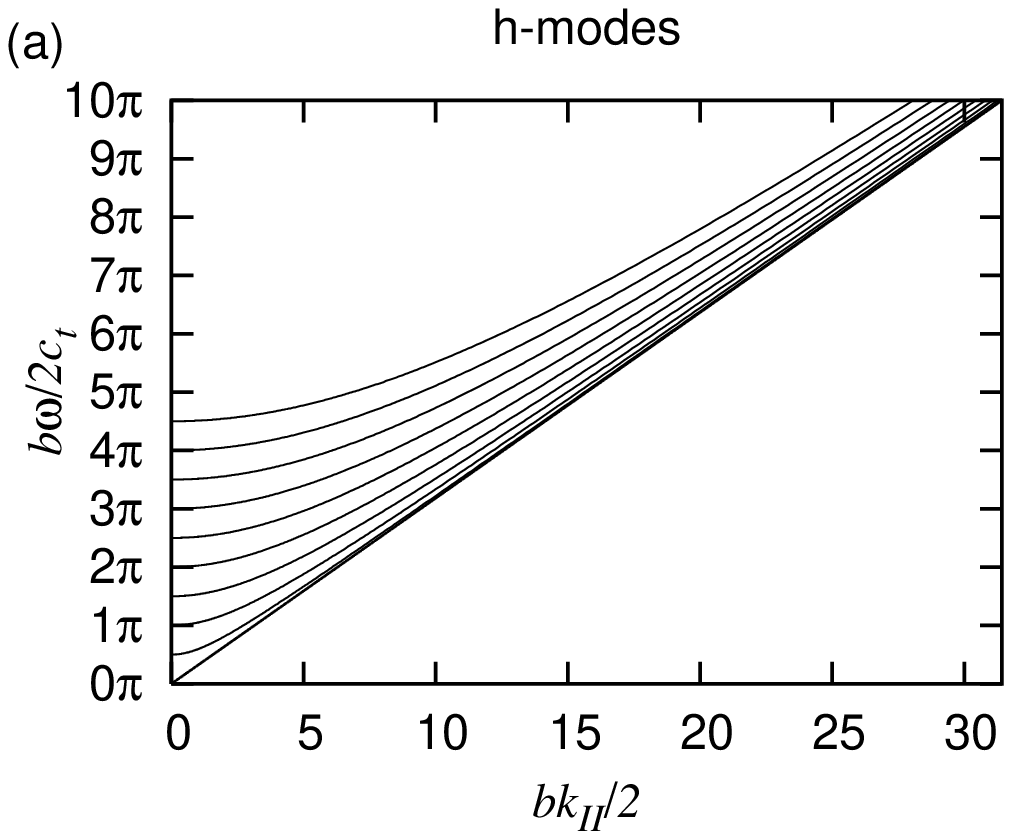}}
\put(25,35){\includegraphics[width=40mm,angle=-90]{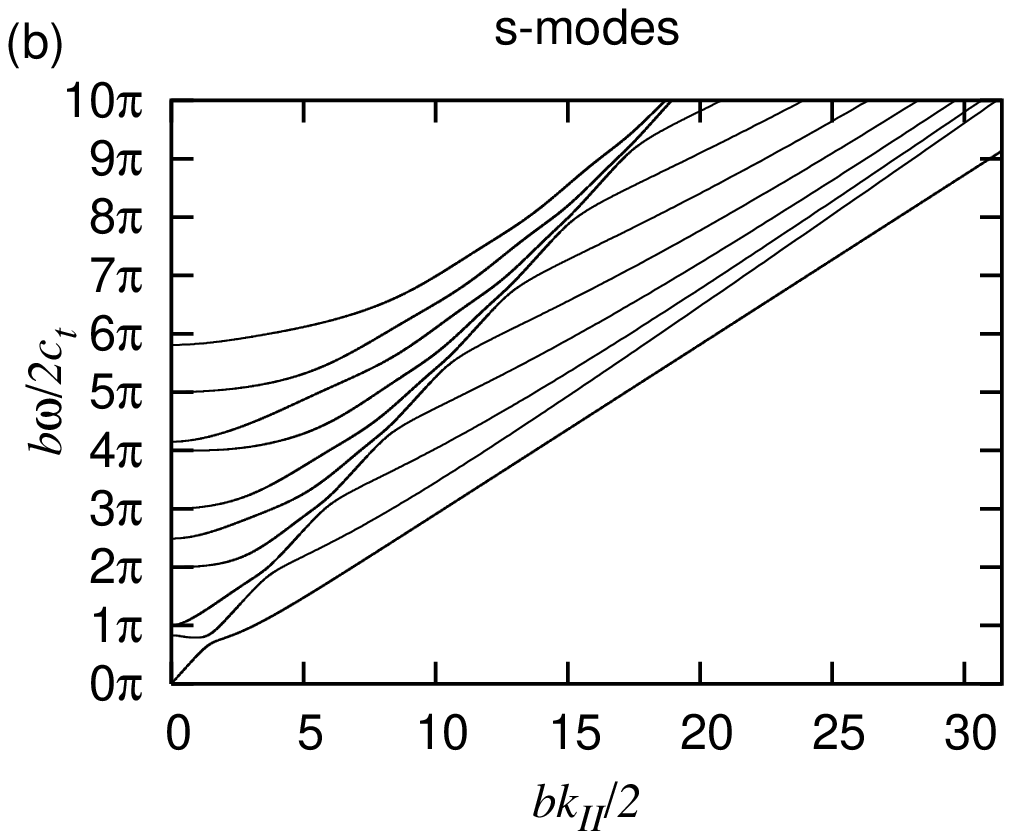}}
\put(70,35){\includegraphics[width=40mm,angle=-90]{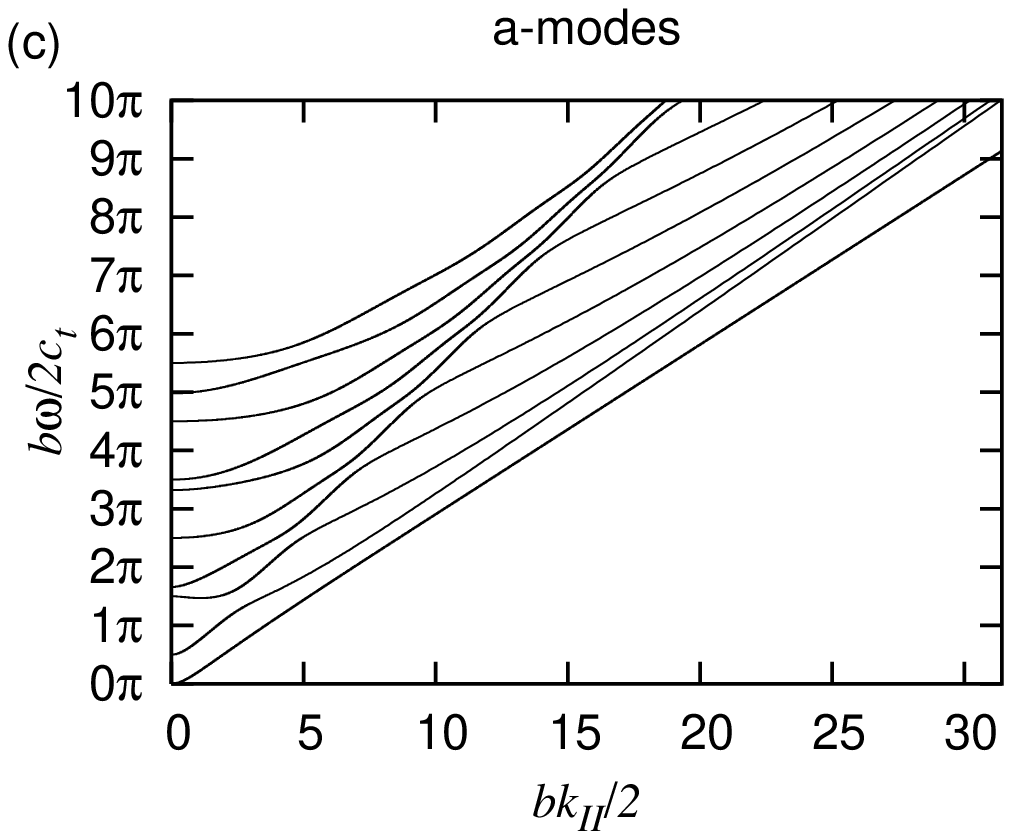}}
\end{picture}
\caption{The branches of the dispersion relations of the phonon eigenmodes
of a free standing thin membrane. ({\em a}) $h$-modes, ({\em b})
$s$-modes and ({\em c}) $a$-modes}\label{fig_disp_rel}
\end{figure*}

When however the system is restricted to finite size, the boundaries of the
system lead to a coupling of longitudinal and transversal modes. 
This coupling is due to the boundary condition at a free
surface: the total stress should vanish.
If the wave incident on the surface is polarized along the $y$
direction  (see Fig.~\ref{reflections} a), the free boundary condition
is satisfied if  the reflected wave has the same polarization. Such a
wave is called a  horizontal shear wave ($h$-wave). The $h$-waves do not
mix at the boundaries with  waves of different polarizations and they form the
typical ``box eigenmodes'',  with the dispersion relation
\begin{equation}
\frac{\omega_{h,n}^2}{c_t^2} = \left(\frac{n\pi}{b}\right)^2 
+ k_\parallel^2 \,. \label{eqn_h_disp}
\end{equation}
Here $k_\parallel$ is the component of the wave vector parallel to the 
membrane surface and $b$ is the thickness of the membrane.
On the other hand, if the polarization of the incident wave is either 
longitudinal or transversal, but in the plane $xz$, then the reflected 
wave will always be a superposition 
of longitudinally and transversally polarized waves. These two waves 
have different propagation velocities, and 
the condition to get eigenmodes is that the plane waves
reconstruct each other after two reflections. The longitudinal and
transversal components of the eigenmode have the same wave vector 
component $k_\parallel$ along the membrane surfaces, but different 
components perpendicular to the surfaces, 
which we shall call $k_\perp^l$ and $k_\perp^t$, respectively. 
The frequency of the eigenmode is then
$\omega^2=c_t^2(k^t)^2=c_l^2(k^l)^2$. These equations give a relation 
between the components $k_\perp^l$ and $k_\perp^t$. 
The eigenmodes obtained in this way fall into two classes: symmetric 
(s-wave) and antisymmetric (a-wave), according to the symmetry of the 
velocity field of the wave with respect to the plane $z=0$. 
The dispersion relation of the $s$- and $a$-waves are given by the equations 
\begin{equation}
\frac{\tan(\frac{b}{2}k_\perp^t)}{\tan(\frac{b}{2}k_\perp^l)}=
-\frac{4k_\perp^lk_\perp^tk_\parallel^2}{[(k_\perp^t)^2-k_\parallel^2]^2}
\label{eqn_s_disp}
\end{equation}
and
\begin{equation}
\frac{\tan(\frac{b}{2}k_\perp^l)}{\tan(\frac{b}{2}k_\perp^t)}=
-\frac{4k_\perp^lk_\perp^tk_\parallel^2}{[(k_\perp^t)^2-k_\parallel^2]^2}\,,
\label{eqn_a_disp}
\end{equation}
respectively. The equations (\ref{eqn_s_disp}) and (\ref{eqn_a_disp}), 
together with $\omega^2=c_t^2(k^t)^2=c_l^2(k^l)^2$ form a set of 
transcendental equations, which can be solved only numerically. The branches 
of each dispersion relation are shown in Fig.~\ref{fig_disp_rel}.
As can be seen, the $h$-modes are just usual ``box modes''. 
Without mixing at the membrane surfaces, all the
modes would be of this type and the results would be identical to the 
ones of Refs.~\cite{dragos1,dragos2}. However, due to the coupling of the
longitudinal and transversal modes, the dispersion
relations of the $a$- and $s$-modes show some interesting properties.

At $k_\parallel = 0$, all the excited branches satisfy the relation 
$\partial \omega/\partial k_\parallel=0$, but for the $s$- and $a$-modes, unlike 
for the $h$-modes, $\partial^2 \omega/\partial k^2_\parallel$ may be either 
positive or negative. If 
$\partial^2 \omega/\partial k^2_\parallel<0$, 
the dispersion curve has a minimum at some value $k_\parallel>0$. The 
lowest branch of the $h$-modes is a straight line,
\begin{equation} \label{disph0}
\omega_{h,0} = c_t k_\parallel 
\end{equation}
but not for the 
$s$- and $a$-modes. For the $s$-modes, 
$\partial \omega_{s,0}/\partial k_\parallel>0$, 
so the group velocity of long wavelength s-phonons is different from zero. 
On the other hand, for the $a$-modes 
$\partial \omega_{a,0}/\partial k_\parallel=0$ 
and $\partial^2 \omega_{a,0}/\partial k^2_\parallel>0$. Therefore the 
group velocity of the $a$-modes is zero at long wavelength and from this 
point of view the $a$-phonons are similar to massive particles. 

Since the low temperature thermal properties of the membranes are 
determined by the dispersion relations of the lowest branches at long 
wavelenghts, we shall take a closer look at these.

\paragraph{Lowest branch of the symmetric modes:} 
If $(b/2)k_\parallel$ converges to zero, the solutions $(b/2)k^t_\perp$ 
and $(b/2)k^l_\perp$ of Eq. (\ref{eqn_s_disp}), and satisfying 
$c_t^2(k^t)^2=c_l^2(k^l)^2$ approach also zero, in such a way that 
$k^t_\perp$ is real, while $k^l_\perp\equiv i\tkp^l$ is imaginary 
\cite{kuehn}. Then Eq. (\ref{eqn_s_disp}) reduces to 
\begin{eqnarray}
\frac{\tan(\frac{b}{2}k_\perp^t)}{\tanh(\frac{b}{2} \tkp^l)}\approx 
\frac{k_\perp^t}{\tkp^l}&=&\frac{4k_\perp^t \tkp^l k_\parallel^2}{
((k_\perp^t)^2-k_\parallel^2)^2} \,.
\label{symmbranch}
\end{eqnarray}
The solution $k_\perp^t=0$ is unphysical (does not satisfy the boundary 
conditions if plugged into the stress formulae), so the only solution is 
$((k_\perp^t)^2-k_\parallel^2)^2=4(\tkp^l)^2 k_\parallel^2$, 
which yields the dispersion relation 
\begin{equation}
\omega_{s,0}=2\frac{c_t}{c_l}\sqrt{c_l^2-c_t^2}k_\parallel\equiv 
c_sk_\parallel \,. \label{disps0}
\end{equation}
So the dispersion relation is linear in the long wavelength limit. 
The group velocity of the s-modes is $c_s$, which, since 
$0.5\le 1-c_t^2/c_l^2<1$ \cite{kuehn}, takes values between 
$\sqrt{2}c_t$ and $2c_t$. 

\begin{figure}[ht]
\centering
\setlength{\unitlength}{1.2 mm}\begin{picture}(120,50)
\put(0,50){\includegraphics[width=60mm,angle=-90]{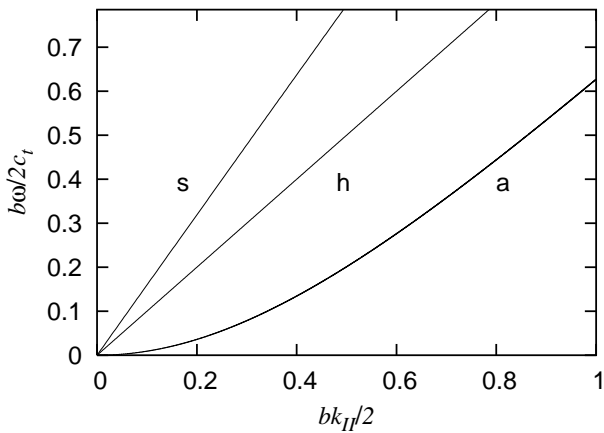}}
\end{picture}
\caption{The three lowest branches of the dispersion relations of the phonon eigenmodes
of a free standing thin membrane shown in the limit of long wavelengths. The linear behaviour
of the $h$- and $s$-branches (labeled {\em h} and {\em s} respectively) as well as the 
quadratic behaviour of the $a$-branch can be seen here.}\label{fig_disp_rel_small}
\end{figure}

\paragraph{Lowest branch of the antisymmetric modes:} 
In the case of antisymmetric modes, if $(b/2)k_\parallel$ converges to zero, 
then both $(b/2)k^t_\perp$ and $(b/2)k^l_\perp$ converge to zero, but 
taking imaginary values: $k^t_\perp\equiv i\tkp^t$ and 
$k^l_\perp\equiv i\tkp^l$, leads to a valid solution \cite{kuehn}. Expanding Eq. (\ref{eqn_a_disp}) 
and using the equation $\omega^2=c_t^2(k^t)^2=c_l^2(k^l)^2$, 
we get the quadratic dispersion relation 
\begin{equation} \label{dispa0}
\omega_{a,0} = \frac{\hbar}{2m^*} k_\parallel^2 \, , 
\end{equation}
as for massive particles of ``effective mass'' 
\begin{equation} \label{effmass}
m^* = \hbar\left[2c_tb\sqrt{(c_l^2-c_t^2)/3c_l^2}\right]^{-1} \,.
\end{equation}
A plot of the lowest branches of the $h$-, $s$- and $a$-modes in the long
wavelength limit is shown in Fig.~\ref{fig_disp_rel_small}.

\section{Thermodynamics}

\subsection{Heat capacity}
The qualitative difference between the dispersion relations of the 
$a$-modes, on one hand, and the $h$- and $s$-modes, on the other hand (see 
Eqs. \ref{disph0}, \ref{disps0}, and \ref{dispa0}), have important 
consequences on the thermodynamic properties of the membrane. 
To show this, let us first calculate the heat capacity of the membrane. 
Phonons obey Bose statistics, so the average population of each phonon 
mode is $n(\omega)=1/[\exp(\beta\hbar\omega)-1]$. If we integrate over 
$k_\parallel$ and sum-up the contributions of all the branches, we arrive 
at the expression
\begin{eqnarray}
C_V &=& \frac{A}{k_BT^2 2\pi}\sum_\sigma\sum_{m=0}^{\infty}\int\limits_{0}^{\infty}\!dk_\parallel\frac{k_\parallel(\hbar\omega_{m,\sigma})^2\exp(\beta\hbar\omega_{m,\sigma})}{(\exp(\beta\hbar\omega_{m,\sigma})-1)^2}\,, \nonumber \\
&&\label{CVgen}
\end{eqnarray}
where $\sigma$ represents the $h$-, $s$-, and $a$-modes, while $\sum_m$ is the 
summation over the branches. The frequencies $\omega_{m,\sigma}$
depend also on $k_\parallel$; $A$ is the area of the membrane. 

\begin{figure*}[ht]
\centering
\setlength{\unitlength}{1.2 mm}\begin{picture}(120,30)
\put(-12,68){\includegraphics[width=80mm,angle=-90]{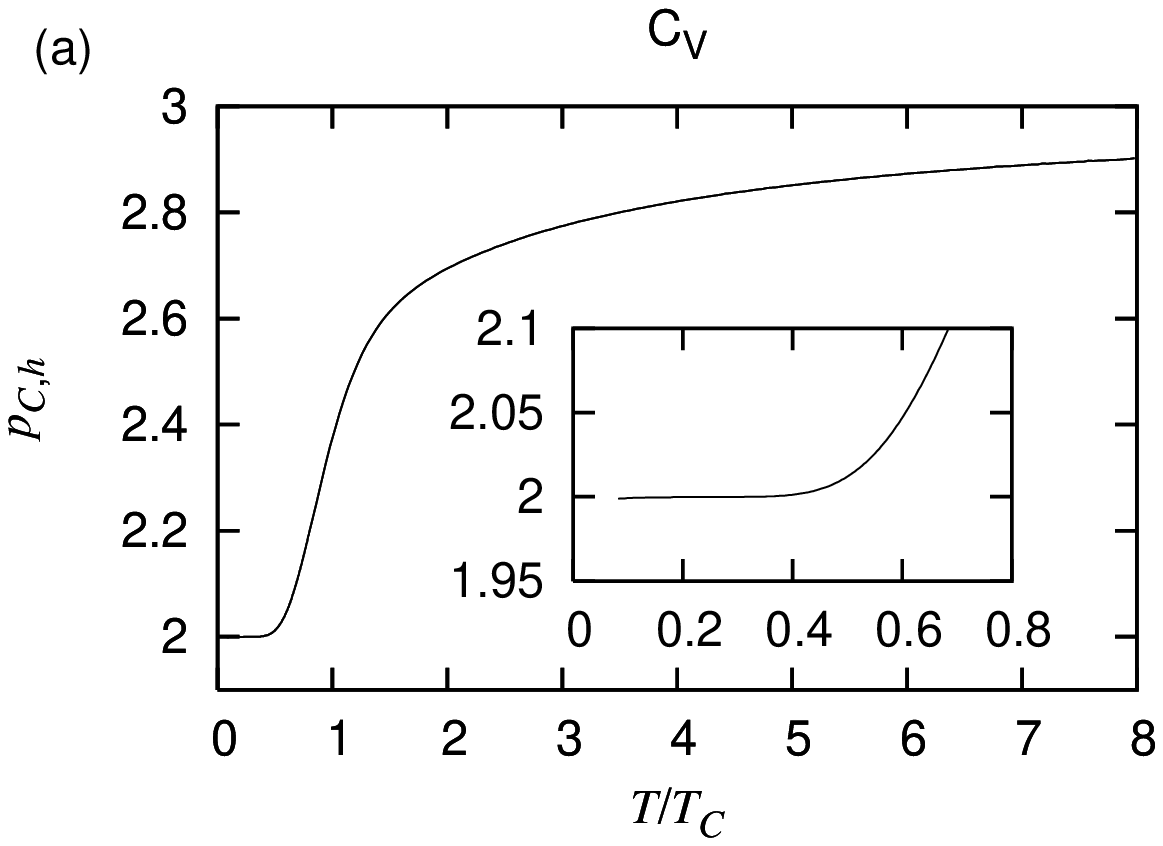}}
\put(35,68){\includegraphics[width=80mm,angle=-90]{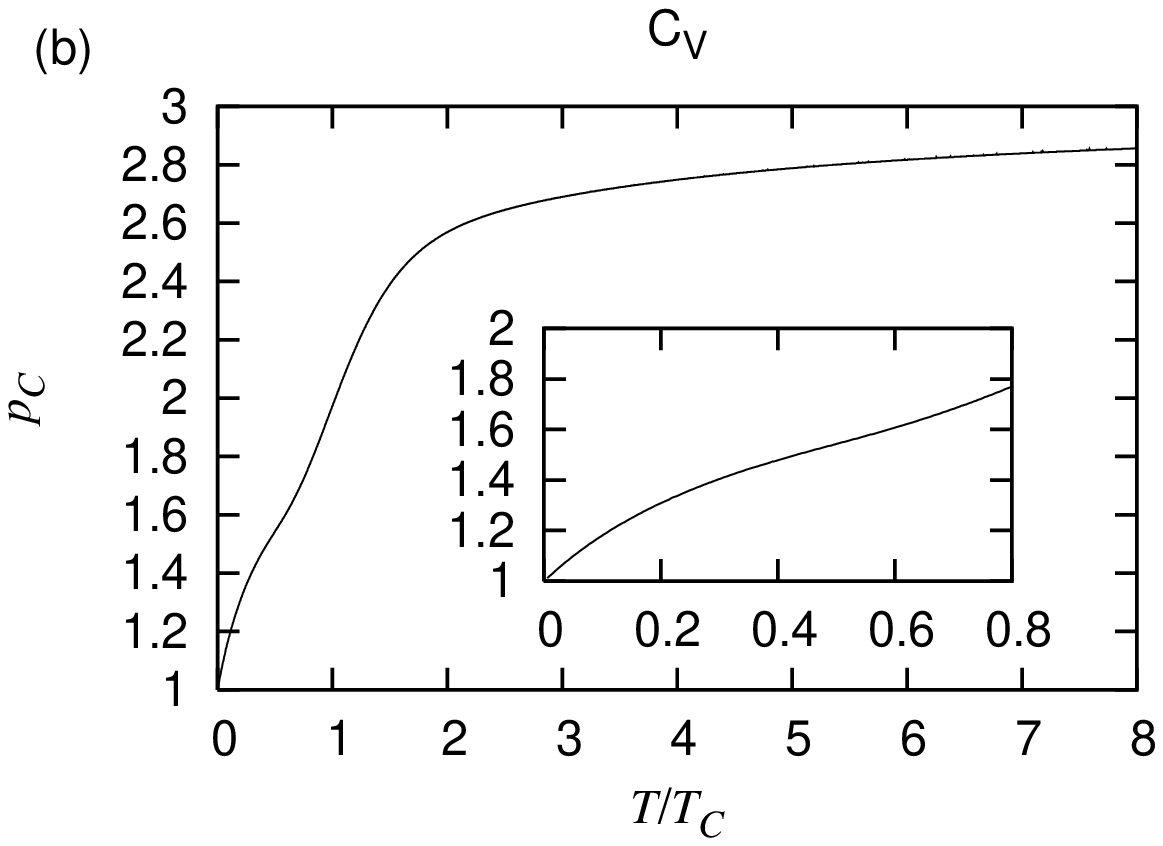}}
\put(82,68){\includegraphics[width=80mm,angle=-90]{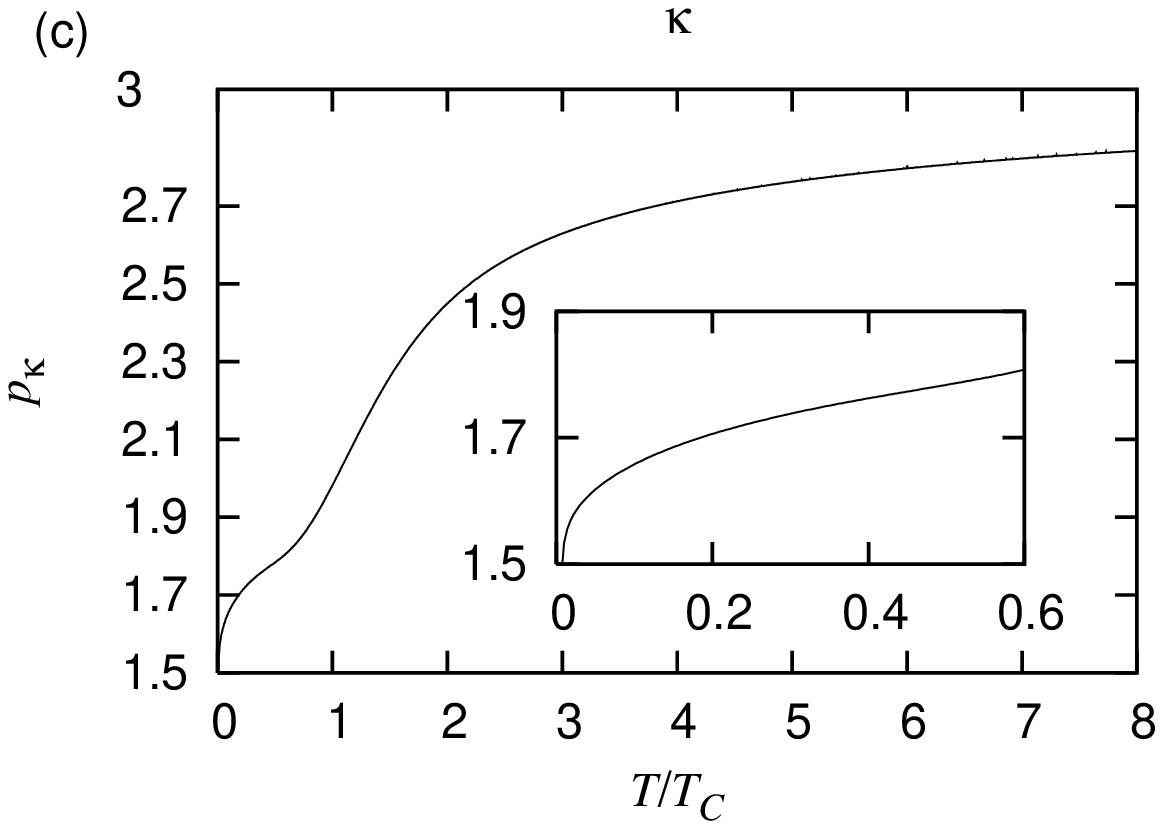}}
\end{picture}
\caption{The temperature exponents $p_f\equiv\partial\ln f/\partial\ln T$ of the heat 
capacity ($f\equiv C$) and heat conductivity ($f\equiv \kappa$) in narrow bridges of thin films. 
The insets show the behaviour of the curves for small values of $T/T_C$.
(a) The temperature exponent of $C_V$, as it would be if the modes were not coupled. 
The dimensionality-crossover around $T_C=(\hbar c_t/2k_B)\cdot b^{-1}$ can be seen here quite nicely. 
For a $100$nm thin membrane the critical temperature of SiN$_x$ is 237 mK.
(b) The temperature exponent of the heat capacity of a thin membrane or a narrow bridge. 
(c) The temperature exponent of the heat conductivity along a narrow bridge.}
\label{fig_cv_kappa}
\end{figure*}

For uncoupled longitudinal and transversal modes, the dispersion laws
have the form $\omega=c_\sigma\sqrt{(n\pi/b)^2+k_\parallel^2}$. In
that case the 3D-to-2D crossover in the phonon gas would manifest itself
through a relatively rapid
change of  the temperature dependence from $T^3$ to $T^2$ at the
temperature  $T_c\equiv \hbar c_t/2bk_B$, as seen in Fig. \ref{fig_cv_kappa} 
(a) \cite{dragos1}.
The corresponding asymptotic temperature dependences of the heat 
capacity, following from Eq.~(\ref{CVgen}), are
\begin{eqnarray*}
C_V\approx\left\{ \begin{array}{ll}
\eta_1T^3,  &T \gg T_c\, ,\\
\eta_2T^2, &   T \ll T_c \, .
\end{array}\right.
\end{eqnarray*}
where
\begin{eqnarray*}
\eta_1=\frac{4\pi Vk_B^4\Gamma(5)\zeta(4) }{(2\pi
  c_3\hbar)^{-3}}\ {\rm and}\ \quad \eta_2=\frac{\pi A k_B^36\zeta(3)}{(2\pi
    c_2\hbar)^{2}}\,.
\end{eqnarray*}
Here, $3/c^3_3\equiv 2/c_t^3+1/c_l^3$, $3/c^2_2\equiv
2/c_t^2+1/c_l^2$,  and $V$ is the volume of the membrane.  The
exponent of the temperature dependence of the heat 
capacity, $p_C\equiv\partial\ln C_V/\partial\ln T$, reflects the 
dimensionality
of the phonon gas distribution: $p_C(T\gg T_c)=3$ and $p_C(T\ll
T_c)=2$ \cite{dragos1,dragos2}.

In the more rigorous case, i.~e. when the dispersion relations are  given by
Eqs. (\ref{disph0}), (\ref{disps0}), and (\ref{dispa0}),  due to the
quadratic dispersion relation of the lowest $a$-mode, we get  a
different temperature behaviour for $T\ll T_c$. Summing only over the
three lowest branches, we get
\begin{eqnarray}
  \label{lowTCV}
  &&C_V \approx Ak_B \left(\alpha T^2 + \beta T \right) \nonumber \\
\alpha& = &\frac{3 \zeta(3) k_B^2}{\pi\hbar^2}\left(\frac{1}{c_t^2}+
    \frac{1}{c_s^2}\right)\, , \
\beta= \frac{\zeta(2)k_B m^*}{\pi\hbar^2}\, .
\end{eqnarray}
(Preliminary estimates show that the contribution of dynamical defects to the 
specific heat are unimportant for such thin films; more detailed calculations 
are in progress.) 
In the high temperature limit, the expected $T^3$ behaviour is obtained, 
but at temperatures around $T_c$, the temperature dependence of $p_C$ is 
somewhat more complicated, converging finally to 1, as $T\to 0$ 
(see Fig. \ref{fig_cv_kappa} b). 

\subsection{Heat conductivity}

A common way to increase thermal insulation of the detector mounted on 
the membrane is 
to cut the membrane so that the central part is connected to
the bulk material only by narrow bridges (see Fig.~\ref{fig_membrane}). 
Therefore, another quantity of interest is 
the thermal conductivity ($\kappa$) along the membrane and bridges. 
Reported widths of the bridges of interest for us are roughly from 
4 ${\rm \mu m}$ upwards \cite{pekola,leivopekola} (We do not discuss here 
the ``dielectric quantum wires'', like in Ref. \cite{DQW}).
As the width of the bridge becomes smaller than the phonon mean free path 
in the uncut membrane, the interaction of phonons with the bridge edges 
should become the main scattering mechanism. 
It is very difficult to solve this problem either analytically or numerically 
for the most general case. 
Instead, we shall try to extract the relevant physical results for our 
problem, using a realistic model. The cutting process leaves the bridge 
edges very rough, so we shall assume that the phonons scatter 
diffusively at the edges, i.e. scattered phonons 
are uniformly distributed over the angles and branches corresponding to 
the same frequency, $\omega$. 

The general expression for the heat current along the rectangular bridge of 
total length $l$ is 
\begin{eqnarray}
\dot Q=
-\frac{1}{l}\sum_{\sigma,m,\bk_\parallel}\hbar\omega(k_\parallel)\tau(k_\parallel)v_x^2(\bk_\parallel)
\frac{\partial n(\omega)}{\partial T}\cdot\frac{\partial T}{\partial x}\, , \label{heatcond}
\end{eqnarray}
where $\tau(\bk_\parallel)$ is the (average) scattering time of a
phonon  having  wavevector $\bk_\parallel$ parallel to the
membrane surface 
and belonging  to the branch $(\sigma,m)$. To simplify the
writing, the dependence  on $\sigma$ and $m$ of the quantities in
Eq. (\ref{heatcond}) was made  implicit. The bridge lies along the
$x$ direction, the $z$ direction is  perpendicular to the
membrane.  We denoted by
$v_{\sigma,n}(\bk_\parallel)\equiv\partial\omega_{\sigma,n}/\partial
k_\parallel$ the group velocity of the phonons. We also assumed
that $\partial T/\partial x$ is not too large, so that the linear
approximation, $\dot{Q} \propto \partial T/\partial x$, holds. Let
us now denote the scattering time of the phonons in the {\em
uncut} membrane by $\tau_{M,\sigma,m}(\bk_\parallel)$, while the
scattering time at the bridge edges is
$\tau_{E,\sigma,m}(\bk_\parallel)$. The effective scattering
time,$\tau_{\sigma,m}(\bk_\parallel)$, is then
\begin{eqnarray*}
\tau^{-1}_{\sigma,m}(\bk_\parallel)=\tau^{-1}_{M,\sigma,m}
(\bk_\parallel)+\tau^{-1}_{E,\sigma,m}(\bk_\parallel)\, .
\end{eqnarray*}
Under the assumption of diffusive scattering at the bridge edges,
$\tau_{E,\sigma,m}(\bk_\parallel)$ can be estimated as
$\tau_{E,\sigma,m}(\bk_\parallel)=w/(v_{\sigma,n}(\bk_\parallel)
\sin\vartheta)$,
where $\vartheta$ is the angle between $k_\parallel$ and the
$x$-direction,  and $w$ is the bridge width.  If we transform the
summation over $\bk_\parallel$ in Eq. (\ref{heatcond})  into an
integral and write $\dot Q\equiv -\kappa\cdot\partial T/\partial
x$, we get
\begin{eqnarray}
\kappa &=& \!\frac{w}{2\pi}\sum_{\sigma,m}\int\limits_0^{2\pi}\!d\vartheta\!\int\limits_0^\infty 
\!dk_\parallel\,k_\parallel \hbar\omega\frac{v^2\cos^2(\vartheta)}{\frac{v|\sin(\vartheta)|}{w}+
\frac{1}{\tau_M}}\frac{\partial n}{\partial T} \nonumber \\
&=& \!\frac{w^2}{2\pi}\sum_{\sigma,m}\int\limits_0^\infty \!dk_\parallel\,k_\parallel 
\hbar\omega\frac{\partial n}{\partial T} v
\!\int\limits_0^{2\pi}\!d\vartheta\frac{\cos^2(\vartheta)}{|\sin(\vartheta)|\!+\!
\frac{w}{l_M}} \label{rawkappa}
\end{eqnarray}
where now $\ell_M\equiv \ell_{M,\sigma,m}(\bk_\parallel)=v_{\sigma,m}(\bk_\parallel)\cdot\tau_{M,\sigma,m}(\bk_\parallel)$ 
is the mean free path corresponding to $\tau_{M,\sigma,m}(\bk_\parallel)$.
Denoting $a=w/\ell_M$ we can write the integral over $\vartheta$ as
\begin{eqnarray*}
C(a)&\equiv& 
\int\limits_0^{2\pi}d\vartheta\frac{\cos^2(\vartheta)}{|\sin(\vartheta)|+a} 
= 4\int\limits_0^1\frac{\sqrt{1-x^2}\,dx}{x+a} \\
&=& 4\!\left[\frac{a\pi}{2}-1+\sqrt{1-a^2}\log\frac{1+
\sqrt{1-a^2}}{a} \right]
\end{eqnarray*}
If edge-scattering dominates, i.~e. $w\ll\ell_M$, the integral only depends logarithmically on a
\begin{eqnarray*}
C(k_\parallel)&\approx&4\log\left(\frac{2}{ae}\right)\, .
\end{eqnarray*}
Considering the small bandwidth of occupied phonon modes as well as 
the long mean free paths in the free membranes at low temperatures, 
we can consider $C(a)\approx C=\text{const}$ over the temperature range 
that we work in.

With this effective ``quasi-constant'' mean free path of the
phonons in the  uncut membrane, we can evaluate the heat
conductivity along the bridge. We did this analytically for the
low temperature limit of a  quasi-2D phonon gas and numerically
for higher temperatures. The analytical low temperature limit gives
\begin{eqnarray}
\kappa &=& \frac{Cw^2k_B}{2\pi}\left[6\zeta(3)\left(\frac{1}{c_t}+\frac{1}{c_s}\right)
\left(\frac{k_BT}{\hbar}\right)^2\right. \nonumber \\
&&\left.+\sqrt{\frac{2m^*}{\hbar}}\frac{15\sqrt{\pi}}{8}\zeta\left(\frac{5}{2}\right)\left(\frac{k_BT}
{\hbar}\right)^{\!3/2\,}\right]\, , \label{lowTk}
\end{eqnarray}
so in the limit of low temperatures $\kappa\propto T^{3/2}$. 
The numerical results are shown in Fig. \ref{fig_cv_kappa} (c). 

\begin{figure}[h]
\centering
\setlength{\unitlength}{1.2 mm}\begin{picture}(60,50)
\put(-5,100){\includegraphics[width=120mm,angle=-90]{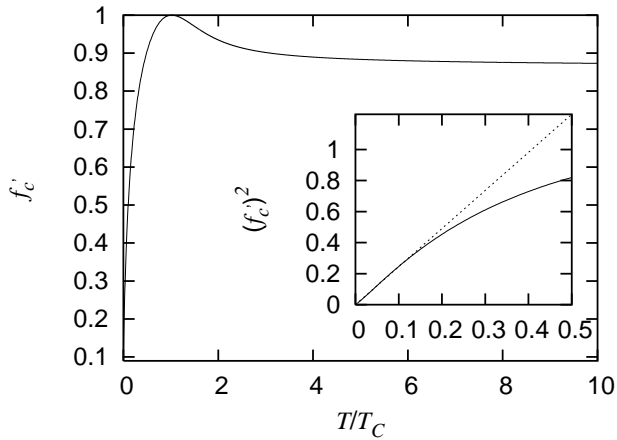}}
\end{picture}
\caption{The cut-off frequency 
of an AC-heated membrane, which is connected to the bulk material by narrow bridges. 
Here we normalized the function to its maximum value, so that
$f_c^\prime=f_c/f_{c,0}\propto\kappa/(lC_V).$
For small values of $T/T_C$ $f_c^2$ becomes linear, which is shown in
the inset.}\label{fig_fc}
\end{figure}

If a membrane, which is connected to the bulk material by narrow
bridges, is heated by an AC current, its thermal cut-off frequency
has the expression $f_c=G/C_V\equiv\kappa/(lC_V)$, where $\kappa$ is the 
thermal
conductivity of the bridge and $C_V$ is the heat capacity of the
membrane.  If all the modes are of the form (\ref{eqn_h_disp})
(uncoupling of the  phonon polarizations at the membrane surfaces),
then both, $C_V$ and  $\kappa$, are proportional to $T^2$ in the
limit of low  temperatures (see \cite{dragos1,dragos2}) and $f_c=\text{const}$.
On the other hand, using Eqs. (\ref{lowTCV}) and (\ref{lowTk}) we obtain 
$f_c\propto T^{1/2}$ in the low temperature limit. The numerical 
results are plotted in Fig. \ref{fig_fc}. 
The increase of $f_c$ with $T$ was observed experimentally in 
Ref.~\cite{leivopekola}, but for a wider temperature range.

\section{Conclusions}

In summary, we used elasticity theory to calculate the  phonon
modes in ultrathin membranes made of homogenous, isotropic silicon
nitride (SiN$_x$). Using the dispersion relations thus obtained,
we calculated the heat capacity and heat conductivity of the
membrane and of bridges, cut out of such membranes. In the low
temperature limit  the phonon gas becomes two-dimensional and
populates three branches of the  dispersion relations, namely the lowest
$h$-, $s$-, and $a$-branches  (see Fig. \ref{fig_disp_rel}). At low temperatures, the
dispersion  relation corresponding to the lowest $a$-branch is
quadratic in $k_\parallel$, while the other two are linear, so at
low temperatures the $a$-branch gives the dominant contribution.
Quite surprisingly, this implies that the universal behavior of
heat capacity in two-dimensional systems is obeyed also by the
phonon gas at low temperatures, where $C_V\propto T$ (see 
Ref.~\cite{universal} and citations therein). 

In the calculation of thermal conductivity along bridges, we
assumed diffusive  scattering of phonons at the bridge edges.
This is justified by the fact  that the cutting process leaves
these edges very rough. If the  width of the bridge decreases
below a certain value (depending on the mean  free path of the
phonons in the uncut membrane) the interaction  with the edges
becomes the main scattering process for the phonons, see
Eq.~(\ref{rawkappa}). In such a case, at low temperatures it  was
found that $\kappa\propto T^{3/2}$ (Eq.~\ref{lowTk}).

If a membrane, connected to the bulk material by narrow bridges,
is heated by an AC current, the amplitude of the temperature
oscillations in the membrane has a cut-off around the frequency
$f_c\equiv G/C_V$. In the low temperature limit, this cut-off
frequency shows an increase with the temperature, as $T^{1/2}$.
Preliminary experimenatal results show an increase of $f_c$ with $T$, but 
on a temperature range much wider than the one in Fig. \ref{fig_fc}. 
This seems to suggest that at higher temperatures other processes have 
to be taken into account in the calculation of thermal characteristics of 
SiN$_x$ membranes.

\section{Acknowledgements}

We thank Prof. J. Lothe for various discussions about elasticity theory and Dr. I. M. Maasilta for useful comments on the manuscript.
We also acknowledge the financial support from the Academy of Finland under the Finnish Center of Excellence Program 2000-2005 (Project No. 44875, Nuclear and Condensed Matter Physics Program at JYFL) and the NorFA Networks, ``Mesoscopic Solid State and Molecular Electronics'' and ``Quantum Transport in Nanoscale systems''.

%

\end{document}